\documentclass{article}
\usepackage{ijcai19}

\usepackage{times}
\usepackage{soul}
\usepackage{url}
\usepackage[utf8]{inputenc}
\usepackage[small]{caption}
\usepackage{graphicx}
\usepackage{amsmath}
\usepackage{booktabs}
\usepackage{algorithm}
\usepackage{algorithmic}
\usepackage{amsmath}
\usepackage{amsthm}
\usepackage{amssymb}
\usepackage{thmtools}
\declaretheorem[starred]{Theorem}

\newtheorem{theorem}{Theorem}[section]

\newtheorem{example}{Example}[section]

\newtheorem{lemma}[theorem]{Lemma}
\newtheorem{proposition}[theorem]{Proposition}

\newtheorem{claim}[theorem]{Claim}
\newtheorem{definition}[theorem]{Definition}

\newcommand{\Halmos}{}

\newcommand{\citep}[1]{\cite{#1}}
\newcommand{\citet}[1]{\citeauthor{#1}~\shortcite{#1}}

\usepackage{csquotes}

\newcommand{\alloc}{\mathcal{S}}

\newcommand{\ADD}{\operatorname{ADDITIVE}}

\newcommand{\LEX}{\operatorname{LEXICOGRAPHIC}}  
\newcommand{\RSPN}{\operatorname{RESPONSIVE}}
\newcommand{\GEN}{\operatorname{GENERAL}}

\newcommand{\SATATTOP}{\operatorname{SATIATED-AT-TOP}}
\newcommand{\SUBMODULAR}{\operatorname{SUBMODULAR}}

\newcommand{\ignore}[1]{}

\title{Competitive Equilibrium with Generic Budgets: Beyond
	Additive}
\author{Moshe Babaioff\thanks{Microsoft Research, \url{moshe@microsoft.com}.} \and
Noam Nisan\thanks{The Hebrew University, \url{noam@cs.huji.ac.il}.} \and 
Inbal Talgam-Cohen\thanks{Technion, \url{italgam@cs.technion.ac.il}.}}
	
\begin{document}
	\maketitle
	\begin{abstract}
		We study competitive equilibrium in the canonical Fisher market model, but with indivisible goods. In this model, every agent has a budget of artificial currency with which to purchase bundles of goods. Equilibrium prices match between demand and supply---at such prices, all agents simultaneously get their favorite within-budget bundle, and the market clears. 
		Unfortunately, a competitive equilibrium may not exist when the goods are indivisible, even in extremely simple markets such as two agents with \emph{exactly the same} budget and a single item. Yet in this example, once the budgets are slightly perturbed---i.e., made \emph{generic}---a competitive equilibrium is guaranteed to exist.         
		In this paper we explore the extent to which generic budgets can guarantee equilibrium existence (and thus related fairness guarantees) in markets with multiple items. We complement our results in \cite{BabaioffNT17} for additive preferences by exploring the case of general monotone preferences, establishing positive results for small numbers of items and mapping the limits of our approach. We then consider \emph{cardinal} preferences, define a hierarchy of such preference classes and establish relations among them, and for some classes prove equilibrium existence under generic budgets.
	\end{abstract}

	\maketitle
		
	\section{Introduction}
	\label{sec:intro}
	We study the notion of competitive equilibrium in markets of indivisible goods without money, in which players have possibly very different budgets of artificial currency.
Such markets capture real-life allocations such as courses to students, shifts to workers, scientific resources within a university, and even food donations among food banks. 
A competitive equilibrium is a fundamental concept in microeconomics, formalizing a steady-state of the market in which prices balance between the agents' demand for the goods, and the supply of goods on the market. 
Such an equilibrium has recently been shown to imply attractive fairness guarantees among players with \emph{different entitlements} for the indivisible goods \cite{BabaioffNT17}.
Our goal is to identify conditions for equilibrium existence by considering \emph{generic} budgets. 

\subsubsection{Motivation 1: Equilibrium existence.}
A fundamental achievement of Arrow and Debreu's \citeyear{AD54} general equilibrium theory in microeconomics  is establishing the existence of a competitive equilibrium under minimal conditions on the market. In an equilibrium, goods are assigned prices, each agent takes her preferred set of goods among the sets that are within her budget, and the market clears. The resulting allocation is Pareto efficient by the \emph{first welfare theorem}. 

Once there are indivisible goods on the market, equilibrium existence is no longer guaranteed.
In this paper we focus on the simple Fisher market model~\citep{BS05} with $m$ goods and $n$ agents (buyers) endowed with positive amounts of artificial currency in the form of budgets $b_1\ge \dots \ge b_n$ (the artificial currency has no use outside the market). 
Each agent has a preference order among all possible bundles of goods.  
The only divergence from the classic model is that goods are indivisible.
To see that an equilibrium need not exist, consider a single indivisible item sold to two agents with the same budget of~1. If the item's price is at most~1, both agents desire it (note that since there is no value for money, an agent is not satisfied if the price is $1$ but she doesn't get the item). If the price is strictly above~1, neither agent can afford it and the market doesn't clear.

Our starting point is that this simple nonexistence example is a knife-edge phenomenon: if the budgets are $1+\epsilon$ and~$1$ for any $\epsilon>0$ instead of precisely equal, an equilibrium does exist (by setting the item price to be in between the two budgets). 
To avoid this knife-edge case, \citet{Bud11} initiated the study of equilibria with \emph{almost}-equal budgets (he also relaxed market clearance and allowed extra copies of certain items, which the current paper does not).
In this paper we take the notion of almost-equal budgets one step further---we consider {\em generic} budgets, i.e., arbitrary budgets (possibly far from equal) to which tiny random perturbations have been added,%
\footnote{A more formal way to define generic budgets is as those which do not satisfy a few simple equalities like $b_1=b_2$ (see Section~\ref{sec:model}; see also the ``Generic property'' page in Wikipedia).}
as a tool to establish existence of competitive equilibrium beyond the single-item case.
While generic budgets do not ensure equilibrium existence in every market model,%
\footnote{For example, in the {\em quasi-linear} model, alongside the indivisible goods there is a single infinitely divisible item that plays the role of money. When budgets are sufficiently large that they are effectively unlimited, this coincides with the combinatorial auction setting, in which market equilibrium existence is guaranteed only for gross substitutes preferences~\citep{GS99}.\label{ftn:quasi}}
in our model there is evidence that genericity can vastly enlarge the set of markets in which an equilibrium exists.
Anecdotal evidence gathered from computer simulations and real-life data suggests that existence in our model is quite common with generic budgets. In particular,
a computerized search we ran over small (two-agent) markets with generic budgets produced \emph{no} non-existence examples;
similarly, a simple t\^atonnement process we applied always found an equilibrium in
larger, real-life markets from the \emph{Spliddit} website \cite{GP14}. 
\emph{This motivates seeking a finer understanding of conditions under which equilibrium existence is guaranteed in Fisher markets with indivisible goods and generic budgets}.

\subsubsection{Motivation 2: Fairness under different entitlements.}  

Which allocations of goods among agents deserve to be called ``fair''? A vast literature is devoted to this question (e.g., \citep{BT96,BCE+16}). Many standard notions of fairness in the literature reflect an underlying assumption that all agents have an equally-strong claim to the goods. However, some agents may be \emph{a priori} much more entitled to the goods than others \cite{RW98}. Examples include partners who own different shares of the partnership's holdings~\citep{CGK87}, different divisions sharing a company's computational resources \citep{GZH+11}, or family members splitting an heirloom \citep{PZ90}.
Despite many applications, there is almost no literature on fair division of indivisibles among agents with different entitlements.%
\footnote{Since in our model utility is non-transferable and money has no intrinsic value, 
	there can be no interpersonal comparison of preference---see Section \ref{sub:intro-setting}. Thus, notions of fairness like 
	maximizing Nash social welfare are inappropriate in our setting.}
Our model perfectly captures these situations, since the entitlement of agent~$i$ can be modeled by her budget $b_i$ (we may assume wlog~that $\sum_i b_i = 1$). 

\citet{BabaioffNT17} generalize to unequal entitlements a classic approach from fair allocation with \emph{equal} entitlements---finding a \emph{competitive equilibrium from equal incomes} (\emph{CEEI}). 
They treat entitlements as budgets, and show that any competitive equilibrium (from unequal incomes) in the resulting Fisher market has compelling fairness guarantees.
In particular, an equilibrium allocation ensures a natural generalization of the \emph{maximin share guarantee} arising from the well-known \emph{cut-and-choose} protocol.%
\footnote{For every agent $i\in [n]$ and any rational number $\ell/d\le b_i$, agent~$i$ is guaranteed to get at least as much as she can by partitioning the items into $d$ bundles and taking the worst $\ell$ out of~$d$.
\citet{FGH+17} independently study a different fairness notion for allocation among unequals, which is unrelated to competitive equilibrium (and in particular not guaranteed to exist for 3~items).
} 
Establishing equilibrium existence thus guarantees the existence of a fair allocation among unequals. On the flip side, non-existence of equilibrium due to indivisibility is to be expected, as indivisibility undermines the ability to fairly allocate. 

If we slightly perturb the budgets representing entitlements to ensure equilibrium existence, the fairness guarantees shown by \citet{BabaioffNT17} change only slightly.
In other words, if we can establish existence of a competitive equilibrium by making the budgets generic, we get almost the same fairness guarantees.
\emph{Thus, a better understanding of competitive equilibrium with generic budgets can help reach fair or approximately fair division of indivisible items among agents with heterogeneous entitlements.}

\subsection{Our model}
\label{sub:intro-setting}

Consider a market setting with $n$ budgeted agents $N$. Unless stated otherwise assume $\sum_i b_i = 1$. Each agent $i$ has an \emph{ordinal}
preference $\preceq_i$ over all subsets of the $m$ indivisible items~$M$. 
We use $S \preceq_i T$ to denote that $i$ \emph{weakly} prefers the set $T$ over the set $S$, and  $S \prec_i T$ to denote that $i$ \emph{strictly} prefers $T$ over $S$, meaning that $S \preceq_i T$ and  $T \npreceq_i S$. When $S \preceq_i T$ and $T \preceq_i S$ we say that $i$ is \emph{indifferent} between $S$ and $T$, and denote this by $S =_i T$.
We assume throughout that every preference $\preceq_i$ is \emph{monotone}, i.e., $S \preceq_i T$ for every $S \subset T$.%
\footnote{Monotonicity is natural in most applications \cite{BCM16} and is the standard assumption in the quasi-linear model. For one exception see \citep{Bud11}.} 
For example, if we are allocating three items $\{A, B, C\}$, a possible preference of agent $i$ might be 
$\emptyset \preceq_i \{A\} \preceq_i \{B\} \preceq_i \{A, B\} \preceq_i \{C\} \preceq_i \{B, C\} \preceq_i \{A, C\} \preceq_i \{A,B,C\}$.
We say that a preference is \emph{strict} if there is never an indifference between sets, 
that is, for every two sets $S\ne T$ it holds that either $S \prec_i T$ or $T\prec_i S$. 

We also consider \emph{cardinal} preferences represented by a valuation function $v_i:2^M\to \mathbb{R}_{\ge 0}$, which assigns to every bundle $S$ a nonnegative value~$v_i(S)$. Wlog, $v_i(\emptyset)=0$. Since money in our model is artificial, it cannot be used as a yardstick against which to measure and compare different agents' preferences, so $v_{i}(S)$ should not be interpreted as an absolute amount by which agent $i$ values bundle $S$. 
We say that an ordinal preference $\succeq_i$ is \emph{represented} by a cardinal valuation $v_i$ if $S \succeq_i T \iff v_i(S) \ge v_i(T)$ and $S \succ_i T \iff v_i(S) > v_i(T)$.

A competitive equilibrium consists of a pair: a price vector $p=(p_1,\dots,p_m)$, and an \emph{allocation} (partition) $\alloc=(\alloc_1,\alloc_2,\dots,\alloc_n)$ of \emph{all} the items among the agents. 
By definition, an allocation is \emph{feasible} (no item is allocated more than once) and \emph{market clearing} (every item is allocated).
Let $p(S)=\sum_{j\in S} p_j$.
We say that $S$ is \emph{within budget~$b_i$} if $p(S)\leq b_i$, and that $S$ is \emph{demanded} by agent $i$ at prices $p$ if $S$ is the most preferred bundle within her budget at these prices (i.e., $p(T) > b_i$ for every $T$ that $i$ strictly prefers to $S$). Using this terminology:

\begin{definition}[CE]
	\label{def:CE}
	A competitive equilibrium (CE) is a pair $(\alloc,p)$ 
	of allocation $\alloc$ and item prices $p$, such that $\alloc_i$ is demanded by agent $i$ at prices $p$ for every $i\in N$.
\end{definition}

(Note that Definition \ref{def:CE} does not include a condition of market clearance as an allocation is always a full partition.)

\subsection{Our results}

In this section we give an overview of our results, using as a reference point the quasi-linear (combinatorial auctions) model for which CE existence is well-studied and well-understood (see Footnote \ref{ftn:quasi} above; see also \cite{BN07}). In contrast, the Fisher model we focus on here has not been studied much, possibly since the nonexistence for single item settings was deemed a nonstarter. 

\subsubsection{General ordinal preferences.}

Our results on the existence of a CE with generic budgets for general preferences can be summarized as follows. 


\begin{Theorem}[Informal]
	A CE with generic budgets always exists 
	in markets with: (i) at most $3$ items and any number of agents; or (ii) $4$ items and $2$ agents.
	On the other hand, there exists a market with 5 items and 2 agents such that for an open interval of budgets,
	no CE exists. 
\end{Theorem}

The above theorem implies that generic budgets can be a useful way of expanding the realm of markets for which a CE exists (recall that without generic budgets a CE is not guaranteed even with $1$ item). The existence results for markets with few items are surprising in comparison to quasi-linear settings, in which a CE may fail to exist even when there are only 2 indivisible items and 2 agents.
However, the theorem rules out a ``sweeping'' existence result that holds for every single market. In Section \ref{sec:general-prefs} we further rule out the possibility that the fundamental \emph{second theorem of welfare} holds for general preferences. This motivates our focus on non-general classes of preferences in the next sections. 

Our theorem leaves open the case of 4 items and more than $2$ agents. In follow-up work, \citet{Seg18} introduces a new technique that settles this open question, showing guaranteed existence for 4 items and 3 agents, and demonstrating possible non-existence for 4 items and 4 agents. 



\subsubsection{Leveled preferences.}
	As we have seen, generic budgets show promise for establishing CE existence, but general preferences turn out to be too general when there are many items. We demonstrate how genericity can guarantee existence of CE for \emph{any} number of items by focusing on a natural subclass of preferences.
	 
	\begin{definition}
		A monotone preference $\succeq$ is \emph{leveled} if for every two bundles $S,T$ such that $|S|>|T|$, $S\succ T$. 
	\end{definition}
	
	That is, an agent with a leveled preference strictly prefers any large bundle to any small one, and among bundles of the same cardinality any preference ordering is feasible. 
	This type of preference fits many real-world settings. One example is two political parties with different numbers of supporting voters (different entitlements) and the allocation of ministerial positions among them \cite{BK04}. The parties often care most about the number of positions they get, and as a secondary consideration about the type of positions.%
	\footnote{As a second example consider the allocation of offices among two departments at a university: each department first cares about the number of offices it receives, and as a secondary consideration about which offices, their precise shapes, etc.}
	This example also nicely demonstrates that generic budgets are often an accurate model for reality, as it is unlikely that the number of supporting voters determining the entitlements will belong to the zero-measure set for which a CE does not exist.

	In Section \ref{sec:leveled} we establish our main existence result for the class of leveled preferences:	
	\begin{Theorem}[Informal]
		In markets with $2$ agents, leveled preferences and arbitrary budgets, for any small enough perturbation of the budgets, a competitive equilibrium
		exists. 
	\end{Theorem}

\subsubsection{Other classes of preferences.}


Looking forward, in order for future work to map the limits of the generic budget approach for establishing CE existence, it must systematically explore relevant preference classes. In combinatorial auctions, a similar exploration of a \emph{hierarchy} of valuation classes, developed in seminal work by \citet{LLN06}, has been the central driving force of the literature for the past decade. The hierarchy in the quasi-linear model consists of additive, substitutes, submodular, XOS, subadditive, limited-complements and finally general valuations (with strict inclusion among these classes). There is typically a sharp dichotomy between limited-complements and fully-general valuations in terms of equilibrium existence etc.

With this in mind, our final goal is to establish a parallel hierarchy of preference (rather then valuation) classes, including well-studied classes from the literature \cite{BCE+16}.
Surprisingly, our main finding is that the second-highest class in the hierarchy before fully-general preferences is preferences representable by submodular valuations! (In the quasi-linear world submodular valuations are relatively low in the hierarchy.) We conclude that in the no-money model, submodular preferences already represent a large chunk of the most relevant preferences for positive results, including all \emph{strict} preferences and even preferences that correspond to valuations with complements.

We now describe the hierarchy: For every type of preference (like additive) we denote in capitals its corresponding class. $\GEN$ is the class of all monotone ordinal preferences. A preference is \emph{submodular} if it can be represented by a submodular valuation $v$: for every two bundles $S,T$, $v(S)+v(T)\ge v(S\cup T)+v(S\cap T)$. A preference is \emph{lexicographic} if there exists a strict preference $\succ^*$ over items such that 
$S\succ T \iff \max_{\succ^*}\{s\in S\setminus T\} \succ^* \max_{\succ^*}\{t\in T\setminus S\}$. I.e., the most preferred item in $S\setminus T$ is preferred over the most preferred item in $T\setminus S$ (this trivially holds if $T\setminus S$ is empty).
Intuitively, an agent with a lexicographic preference would trade all of her items for a single higher-ranked item.
The remaining definitions as well as the proof establishing the following strict hierarchy are deferred to Section \ref{sec:classes-of-prefs}:
\begin{eqnarray*}
&\LEX \subsetneq \ADD& \nonumber\\
&\subsetneq \RSPN \subsetneq \SUBMODULAR \subsetneq \GEN.& \label{eq:hierarchy}
\end{eqnarray*}
(We remark that the class of leveled preferences is incomparable with the classes in the hierarchy below $\GEN$, in the sense that none of these classes contains the other.)

To state our result for $\SUBMODULAR$, we need the following definition (see Section \ref{sec:classes-of-prefs}). We say that a monotone preference $\preceq_i$ satisfies \emph{satiation-only-at-the-top} (and belongs to class $\SATATTOP$) provided that if item $j$ adds nothing to a given set $T$, it also adds nothing to any superset of $T$. This is quite a weak condition, satisfied in particular by any strict monotone preference. We show:
\begin{Theorem}[Informal]
	The class $\SUBMODULAR$ is equal to $\SATATTOP$.
\end{Theorem}
%
The above thus gives a complete characterization of monotone ordinal preferences that can be represented by submodular valuations. 

Returning to the question of generic CE existence, in Section \ref{sec:classes-of-prefs} we establish this for $\LEX$. The class $\ADD$ has been considered in \cite{BabaioffNT17,Seg18}, and the state-of-the-art is positive results for 2 additive agents with additional conditions; removing these conditions (and extending to $\RSPN$ and beyond) is currently the central open question in this line of work. 


\subsection{Additional related work}

Our model is a special case of Arrow-Debreu exchange economies with indivisible items. Several other variants have been studied.
The line of work originating with \citep{Sve83,Mas87} (see \citep{NW16} for a recent example) 
considers markets with indivisibilities in which an infinitely divisible good plays the role of money. 
The line of work originating with \citep{SS74,Sve84} focuses on the house allocation problem. 
Several works assume a continuum of agents like \cite{Mas77}, and/or study relaxed CE notions like 
\citep{RC07}. Closest to ours are the model of \emph{combinatorial assignment} \cite{Bud11}, which allows non-monotonic preferences, and the model of \emph{linear markets} 
\cite{DPS03}, which crucially uses non-generic budgets. 

\citet{Bud11} circumvents the non-existence of a CEEI due to indivisibilities by weakening the equilibrium concept (relaxing market clearance and sometimes requiring more copies of certain items). 
His work focuses exclusively on budgets that are almost equal. In the same model, \citet{OPR16} show PPAD-completeness of computing an approximate CEEI as well as other hardness results. 
The preferences used in the hardness proofs are non-monotone, leading Othman et al.~to suggest restricting the preferences (as we do here) as a way around their negative results. \citet{BHM15} study the existence of exact CEEI for two valuation classes (perfect substitutes and complements) with non-generic budgets.
Recent independent work by \citet{BarmanKV18} uses CEs in Fisher markets as a tool to achieve a different fairness guarantee than ours---(approximate) envy-freeness rather than (generalized) maximin share guarantees---for additive preferences.

\subsection{Organization}
Section \ref{sec:model} contains several preliminaries,
Section \ref{sec:general-prefs} includes our results for general ordinal preferences, 
and Section \ref{sec:leveled} presents our existence result for leveled preferences. 
Section \ref{sec:classes-of-prefs} presents classes of preferences and their hierarchy. 


		
	\section{Preliminaries}
	\label{sec:model}
	In this section we formally define budget-exhausting prices, Pareto optimality and welfare theorems, and generic budgets.

Given budgets $b$ and an allocation $\alloc$, we say that prices $p$ are \emph{budget-exhausting} if $p(\alloc_i)=b_i$ for every agent~$i$. Note that if $p$ is budget-exhausting then every agent is allocated, i.e., $\alloc_i\neq \emptyset$ for every $i$ (since $b_i>0$). 
We observe that budget-exhaustion is wlog when every agent is allocated (this will come in handy in Section \ref{sec:general-prefs}):

\begin{claim}[Budget-exhausting prices are wlog]
	\label{cla:exhaust-wlog}
	For every CE $(\alloc,p)$ such that $S_i\neq \emptyset$ for every $i\in N$, 
	there exists a CE $(\alloc,p')$ in which $p'$ is budget-exhausting.
\end{claim}

\proof 
	As every agent is allocated at least one item, we can raise the price of that item in her allocated bundle until her budget is exhausted. The new prices form a CE with the original allocation since every agent still gets her demanded set. 
	\Halmos
\endproof

The next definition formulates what it means in our model for an allocation (CE or otherwise) to be ``efficient'' (since we are in a non-transferable-utility model (no-money), maximizing welfare i.e.~the sum of agents' values, is irrelevant):

\begin{definition}[PO]
	Consider a market with preferences $\{v_i\}_{i\in N}$. An allocation $\alloc$ is Pareto optimal (a.k.a.~Pareto efficient or PO) if no allocation $\alloc'$ \emph{dominates} $\alloc$, 
	i.e., if for every $\alloc'\ne\alloc$ there exists an agent~$i$ for whom $\alloc_i \succ_i \alloc'_i$.
\end{definition}

PO plays a role in two kinds of fundamental welfare theorems in economics, which apply to various market models and equilibrium notions. The first is known to apply to CEs in our setting:

\begin{theorem}[First welfare theorem]
	\label{thm:first-welfare-thm}
	Let $(\alloc,p)$ be a CE. Then $\alloc$ is PO.
\end{theorem}

To discuss the second welfare theorem we use the following terminology: 
Given a market with preferences $\{\preceq_i\}_{i\in N}$ and budget profile~$b$, an allocation $\alloc$ is \emph{supported} in a CE if there exist item prices $p$ such that $(\alloc,p)$ is a CE. Where only preferences are given, we say $\alloc$ is supported in a CE if there exist prices $p$ and budgets $b$ such that $(\alloc,p)$ is a CE.
The second welfare theorem, where it holds, states that every PO allocation is supported in equilibrium. In Section \ref{sec:general-prefs} we refute this fundamental theorem for the basic Fisher model. I.e., we show there are some socially-efficient allocations which cannot be realized in equilibrium even if the social planner has the power to set the budgets as she desires.


\subsubsection{Generic budgets (versus different budgets).}

We are interested in showing \emph{generic} existence of a CE for classes of markets. 
We use the standard notion of genericity, i.e., ``all except for a zero-measure'', or equivalently, ``with tiny random perturbations''. By generic existence we thus mean that for every market in the class, for every vector of budgets except for a zero-measure subset, a CE exists. 
An equivalent way to say this is: for every market in the class, for every vector of budgets, by adding tiny random perturbations to the budgets we get a new instance in which a CE exists with probability 1.

A useful way of specifying a zero-measure subset of budgets is as those which satisfy some condition, for instance, $b_1=2b_2$. 
The conditions we use differ among different CE existence results, and for concreteness we shall list them explicitly within each result; we emphasize however that the conditions themselves are irrelevant to our contribution, as long as the measure of budgets satisfying them is zero.

Generic budgets are not to be confused with \emph{different} (or \emph{arbitrary}) budgets, by which we mean budgets that are not necessarily equal or almost equal to one another.

	\section{General preferences}
	\label{sec:general-prefs}
	In this section we prove our results for general ordinal monotone preferences. 
We begin with two existence results.

\begin{proposition}[2 or 3 items]
	\label{pro:3-items}
	Consider $n\ge 2$ agents with monotone preferences over $m\le 3$ items, and budget profile $b$. If $b_1>b_2>b_3\geq 0$ (where $b_3=0$ if $n=2$), then a CE exists.
\end{proposition}

\proof
For 2 items, the following is a CE: let agent $1$ pay $b_1$ (the highest budget) for his most preferred item, and let agent $2$ pay $b_2$ (the second-highest budget, strictly lower) for the remaining item. 

For 3 items, first observe that in any CE, an agent whose budget is not among the three highest will remain unallocated. 
We proceed by case analysis:
\begin{itemize}
	\item If $b_1>3 b_2$, then agent $1$ gets all 3 items, each for a price of $b_1/3>b_2$.
	
	\item If $3 b_2\ge b_1 > 2 b_2$, then agent $1$ gets the bundle of size 2 that he most prefers, and pays $b_1/2>b_2$ for each item, and agent $2$ gets the remaining item for price of $b_2$.
	
	\item If $2 b_2\ge b_1 > b_2+b_3$, then if the pair of items that agent $1$ prefers most does not contain agent $2$'s most preferred item, give agent $1$ this pair and charge him $b_1/2$ for each item, and give agent $2$ the remaining item for price $b_2$. Otherwise, give agent $2$ his second most preferred item for a price of $b_2$, give the other 2 items to agent $1$ and charge him $b_2+\epsilon>b_2$ for agent $2$'s most preferred item, and $b_1-b_2-\epsilon>b_3$ for the other item. 
	
	\item If $b_1=b_2+b_3$, then if there is an item that agent $1$ prefers over all pairs of items, he gets the item and pays his budget, while each other agent -- in the order of their budgets -- picks his most preferred item and pays his budget. 
	If there is no such item, agent $2$ gets his most preferred item for price $b_2$, and agent  $1$ get the remaining 2 items, each for a price of $b_1/2<b_2$.
	
	\item If $b_2+b_3 > b_1 >b_2$, then each agent in the order of budgets picks his most preferred item out of the remaining items, paying his budget. 
\end{itemize}
It is not hard to verify that each of these price vectors indeed form a CE for the given budgets. 
\Halmos
\endproof

\begin{proposition}[4 items]
	\label{pro:4-items}
	Consider $n=2$ agents with monotone preferences over $m=4$ items, and budget profile $b$. If $b_1\notin\{4b_2, 3b_2, 3b_2/2\}$, then a CE exists.
\end{proposition}

\proof
Denote the items by $\{A,B,C,D\}$. We partition the space of budgets by the ratio of $b_1$ and $b_2$ as follows:
\begin{itemize}
	\item If $b_1>4b_2$: Price every item at $b_1/4$ and give all items to agent 1. 
	\item If $4b_2>b_1>3b_2$: Give agent 1 the bundle of size $3$ that he most prefers, and price each item at $b_1/3$. Give the leftover item to agent 2 at price $b_2$.  
	\item If $3b_2>b_1>3b_2/2$: See Claim \ref{cla:4-items-case-1}.
	\item If $3b_2>b_1>3b_2/2$: See Claim \ref{cla:4-items-case-2}.
\end{itemize}		

\begin{claim}
	\label{cla:4-items-case-1}
	A CE exists if $3b_2>b_1>3b_2/2$.
\end{claim}

\proof[Proof of Claim \ref{cla:4-items-case-1}.]
If agent 1 prefers any triplet of items to any pair, then give agent 2 the item that he most prefers at price $b_2$, and give agent 1 the remaining triplet at price $b_1/3$ per item. Otherwise assume without loss of generality that $\{A,B\}$ is agent 1's most preferred pair. If agent 1 prefers $\{A,B\}$ only to one triplet, without loss of generality $\{B,C,D\}$, then there are three cases:
\begin{itemize}
	\item Case 1: Agent 2's favorite item is not $A$. Give agent 2 the item that he most prefers at price $b_2$, and give agent 1 the remaining triplet at price $b_1/3$ per item.
	
	\item Case 2: Agent 2 prefers $\{C,D\}$ to $\{B\}$. If $b_1>2b_2$, give agent 1 the pair $\{A,B\}$ and agent 2 the pair $\{C,D\}$. Set the prices to be $(b_1-b_2+\epsilon, b_2-\epsilon, b_2/2, b_2/2)$. Otherwise if $b_1\le 2b_2$ set the prices to be $(b_2+\epsilon, b_1-b_2-\epsilon, b_2/2, b_2/2)$.
	
	\item Case 3: Give agent 1 the triplet $\{A,C,D\}$ and agent 2 the item $B$. Set the prices to be $(b_2+\epsilon, b_2, \frac{b_1-b_2-\epsilon}{2},$  $\frac{b_1-b_2-\epsilon}{2})$. 
\end{itemize}

Assume now agent 1 prefers $\{A,B\}$ to both $\{A,C,D\}$ and $\{B,C,D\}$. If $b_1>2b_2$, then give agent 1 the pair $\{A,B\}$ and give agent 2 the pair $\{C,D\}$. Set the prices to be $(b_1/2,b_1/2,b_2/2,b_2/2)$. Otherwise $2b_2\ge b_1$ and there are two cases:
\begin{itemize}
	\item Case 1: Agent 2 prefers $\{C,D\}$ to $\{A\}$ or to $\{B\}$; assume without loss of generality the latter. Then give agent 1 the pair $\{A,B\}$ and give agent 2 the pair $\{C,D\}$. Set the prices to be $(b_2+\epsilon,b_1-b_2-\epsilon,b_2/2,b_2/2)$.
	\item Case 2: Agent 2 prefers $\{A\}$ and $\{B\}$ to $\{C,D\}$. Assume without loss of generality that agent 1 prefers $\{A,C,D\}$ to $\{B,C,D\}$. Give agent 1 the triplet $\{A,C,D\}$ and give agent 2 the item $B$. Set the prices to be $(b_2+\epsilon, b_2, \frac{b_1-b_2-\epsilon}{2}, \frac{b_1-b_2-\epsilon}{2})$.
\end{itemize}
This completes the proof of Claim \ref{cla:4-items-case-1}.
\endproof

\begin{claim}
	\label{cla:4-items-case-2}
	A CE exists if $3b_2/2>b_1>b_2$.
\end{claim}

\proof[Proof of Claim \ref{cla:4-items-case-2}.]
Without loss of generality let $A$ be the single item most preferred by agent 2. Let $\delta = b_1 - b_2$. 

\begin{itemize}
	\item Case 1: The complement to a pair that agent 1 prefers over $\{B,C,D\}$ appears before $A$ in agent 2's preference ordering. Give agent 1 the most preferred such pair, which must include A by monotonicity, and give agent 2 its complement.
	
	\begin{itemize}
		\item Subcase (i): Agent 1 gets his most preferred pair. Set the prices to be $b_1/2, b_1/2$ for agent 1's items, and $b_2/2, b_2/2$ for agent 2's items.
		
		\item Subcase (ii): Agent 1 gets his second most preferred pair, without loss of generality $\{A,C\}$ where the most preferred pair is $\{A,B\}$. This means that $\{C,D\}$ appears after $\{A\}$ in agent 2's preference ordering. Set the prices to be $(b_1-b_2/2, b_2/2+\epsilon, b_2/2, b_2/2-\epsilon)$, where $\epsilon <\delta$.
		
		\item Subcase (iii): Agent 1 gets his third most preferred pair, without loss of generality $\{A,D\}$ where the first and second most preferred pairs are $\{A,B\}$, $\{A,C\}$. This means that $\{C,D\}$ and $\{B,D\}$ appear after $\{A\}$ in agent 2's preference ordering. Set the prices to be $(b_1/2+\delta, b_2/2, b_2/2, b_1/2-\delta)$.
	\end{itemize}
	
	\item Case 2: agent 1 prefers $\{A\}$ over $\{B,C,D\}$. Give agent 1 the item $A$ and give agent 2 the bundle $\{B,C,D\}$. Set the prices to be $(b_1, b_2/3, b_2/3, b_2/3)$.
	
	\item Case 3: Give agent 1 the bundle $\{B,C,D\}$ and give agent 2 the item $A$.
	
	\begin{itemize}
		\item Subcase (i): Agent 1 has one pair preferred over $\{B,C,D\}$, without loss of generality $\{A,B\}$. This means that $\{C,D\}$ appears after $\{A\}$ in agent 2's preference ordering. Set the prices to be $(b_2, b_1-2\epsilon, \epsilon, \epsilon)$, where $\delta/2 < \epsilon < \delta$.
		
		\item Subcase (ii): Agent 1 has two pairs preferred over $\{B,C,D\}$, without loss of generality $\{A,B\}$ and $\{A,C\}$. This means that $\{C,D\}$ and $\{B,D\}$ appear after $\{A\}$ in agent 2's preference ordering. Set the prices to be $(b_2, \frac{b_1-\epsilon}{2}, \frac{b_1-\epsilon}{2}, \epsilon)$, where $\epsilon < \delta$.
		
		\item Subcase (iii): Agent 1 has three pairs preferred over $\{B,C,D\}$, these pairs are $\{A,B\}$, $\{A,C\}$ and $\{A,D\}$. This means that $\{C,D\}$, $\{B,D\}$ and $\{B,C\}$ appear after $\{A\}$ in agent 2's preference ordering. Set the prices to be $(b_2, b_1/3, b_1/3, b_1/3)$.
	\end{itemize}
\end{itemize}
This completes the proof of Claim \ref{cla:4-items-case-2}.
\endproof
Combining Claim \ref{cla:4-items-case-1} with Claim \ref{cla:4-items-case-2}, Proposition \ref{pro:4-items} follows.
\endproof

We now switch to negative (i.e.~non-existence) results, which we prove via the next example. The example is inspired by a classic CE non-existence result in the completely different model of quasi-linear utilities (combinatorial auctions), where one agent (Alice) treats two items as complements and another (Bob) as substitutes. The three extra items help adapt the example to our no-money model. 

\begin{example}[2-player 5-item market with no CE] 
	\label{ex:2-players-5-items}
	Let agent 1 be Alice and let agent 2 be Bob. Denote the items by $A,B,C,D,E$. The ordinal preferences are induced by cardinal values as follows.
	For the first two items, Alice and Bob have different perspectives. Alice values item $A$ at $10$ and item $B$ at $20$, and the pair at $700$. Bob has a value of $500$ for $A$ and $501$ for $B$, and a value of $502$ for both. Both Alice and Bob have additive preferences over the three items $C,D$ and $E$, valuing them at $201,202$ and $203$, respectively.  
	For both agents, the value of a set is additive across the pair $\{A,B\}$ and the three remaining items; for example, if Alice gets $\{A,B,C,D\}$ her value is $700+201+202=1103$. 
\end{example}
 

\begin{theorem}[5 items]
	\label{thm:no-CE-5-items}
	There exist monotone strict preferences for $2$ agents over $5$ items such that for any budget profile $b$ satisfying $(4/3)b_2>b_1>b_2$, no CE exists. 
\end{theorem}

\begin{theorem}[No second welfare theorem]
	\label{thm:no-2nd-welfare}
	There exist monotone strict preferences for $2$ agents over $5$ items, and a PO allocation $\alloc$, such that $\alloc$ is not supported in a CE for any budget profile $b$.  
\end{theorem}	


The proofs of Theorems \ref{thm:no-CE-5-items} and \ref{thm:no-2nd-welfare} rely on the following key lemma (Theorem \ref{thm:no-2nd-welfare} follows from it immediately):

\begin{lemma}
	\label{lem:PO-not-supported}
	Let $\alloc$ be any allocation in Example \ref{ex:2-players-5-items} in which
	Alice gets an item $X_1\in \{A,B\}$ and items $Y_1,Y_2\in \{C,D,E\},Y_1\neq Y_2$, and Bob gets the remaining items
	$X_2\in \{A,B\}$ and $Y_3\in \{C,D,E\}$. 
	Then $\alloc$ is PO, and not supported in a CE for any budget profile $b$.
\end{lemma}

\proof
Allocation $\alloc$ is PO since both Alice and Bob prefer $B$ over $A$, and both have the same preference order over items $C$, $D$ and $E$. Thus there is no exchange of items that improves the situation for both agents simultaneously. Assume for contradiction that $\alloc$ is supported in a CE.
Since both Alice and Bob prefer Alice's bundle $\{X_1,Y_1,Y_2 \}$ over Bob's bundle $\{X_2,Y_3\}$, to support $\alloc$ in a CE it must be the case that $b_1>b_2$. 
By Claim \ref{cla:exhaust-wlog} we may assume both agents exhaust their budgets, and so $b_1= p_{X_1} + p_{Y_1}+p_{Y_2}$ and $b_2= p_{X_2} + p_{Y_3}$. 	
In such a CE, since Alice prefers replacing $X_1$ by $Y_3$ and replacing $\{Y_1,Y_2\}$ by $X_2$, it must be that $p_{X_1}<p_{Y_3}$ and $p_{Y_1}+p_{Y_2}<p_{X_2}$. This implies that 
$b_1= p_{X_1} + p_{Y_1}+p_{Y_2}< p_{X_2} + p_{Y_3}=b_2$, a contradiction.
\Halmos
\endproof


\proof[{Proof of Theorem \ref{thm:no-CE-5-items}}]
By the first welfare theorem (Theorem \ref{thm:first-welfare-thm}), to show that no CE exists we only need to consider PO allocations. By strict monotonicity, every item is allocated. 
By Claim \ref{cla:exhaust-wlog} we can further assume that if there were a CE, then both players would exhaust their budgets. 
We begin by observing that since Alice has more money, it is not possible that Bob gets both $A$ and $B$ -- Alice can afford Bob's bundle, and she prefers $\{A,B\}$ over $\{C,D,E\}$. 
On the other hand, Alice cannot get $\{A,B\}$ or any superset of it. This is because in such a CE one of the items $A$ or $B$ will have price at most $b_1/2<b_2$, and Bob prefers that item over any pair of items from his set, and there is such a pair that costs at least $(2/3)b_2> b_1/2$, thus Bob will deviate.   
We are left to consider the case that Alice gets one of $A$ and $B$, and Bob gets the other. If Alice also gets all three items in $\{C,D,E\}$, the allocation is not PO as it is dominated by giving Alice the set $\{A,B\}$ and Bob the rest.  
If Alice gets at most one of the items in $\{C,D,E\}$, she would rather exchange bundles with Bob, and can afford it, so it is not a CE. 
If Alice gets one of $A$ or $B$, and two items from the set $\{C,D,E\}$, Lemma \ref{lem:PO-not-supported} rules out the existence of a CE, completing the proof.
\Halmos
\endproof
				
	\section{Leveled preferences} 
	\label{sec:leveled}
	In this section we prove that leveled preferences ensure generic CE existence for 2 agents. 
This is in contrast to the quasilinear (combinatorial auctions) model.%
\footnote{If agent 1 has values $(2, 2, 2)$ for bundles $(A,B,AB)$ over 2 items, and agent 2 has values $(0, 0, 3)$ for these bundles, then the valuations represent leveled preferences but one can verify that with quasilinear utilities a CE does not exist.}

\begin{theorem}
	Consider $n=2$ agents with leveled preferences over $m$ items, and budget profile $b$. If $mb_1$ is not an integer then a CE exists.
\end{theorem}

\proof
	Let $p^*=1/m$, $k_1=\lfloor m b_1\rfloor$ and $k_2=\lfloor m b_2 \rfloor$. Intuitively, $p^*$ is the ``fair'' price for an item, and $k_1,k_2$ are the target number of items to allocate to each agent in order to achieve \emph{budget-proportionality}.
	By assumption, $m\cdot b_1$ is not an integer and so $k_1+k_2=m-1$. By definition, $b_1/k_1 > p^* > b_1/(k_1+1)$, and $b_2/k_2 > p^* > b_2/(k_2+1)$. 
	
	Assume wlog that  $b_1/(k_1+1) > b_2/(k_2+1)$ (the analysis of the complementary case is similar and is omitted for space considerations). Give agent 2 her most preferred set of $k_2$ items, and agent 1 the remaining $k_1+1$ items. Price each of agent 1's items at $b_1/(k_1+1)$, and each of agent 2's items at $b_2/k_2$. 
	
	We argue that no agent wishes to deviate: Agent 1 would only want to deviate to a bundle with $\ge k_1+1$ items, but all such bundles are above her budget---she is currently exhausting her budget with items priced $b_1/(k_1+1)$, and all other items cost $b_2/k_2 > p^* > b_1/(k_1+1)$. Agent 2 gets her most preferred set of $k_2$ items, so would only want to deviate to a bundle with $>k_2$ items, but all such bundles are above her budget -- item prices are at least $b_1/(k_1+1) > b_2/(k_2+1)$.
	\Halmos
\endproof

	\section{A hierarchy of preferences}
	\label{sec:classes-of-prefs}
	In this section we explore the hierarchy of preference classes presented in Section \ref{sec:intro}: 
\begin{eqnarray}
&\LEX \subsetneq \ADD& \label{eq:hierarchy}\\
&\subsetneq \RSPN \subsetneq \SUBMODULAR \subsetneq \GEN.\nonumber
\end{eqnarray}
We first fill in the missing definitions and prove the hierarchical relations in \eqref{eq:hierarchy}.
We then characterize the hierarchy's next-to-final level and show generic CE existence for its first level (in fact, the same argument establishes existence for unit-demand valuations).

$\ADD$ is the class of preferences that can be represented by an additive valuation, where a valuation $v$ is \emph{additive} if $v(S)=\sum_{j \in S}v(\{j\})$ for every bundle~$S$. $\RSPN$ is the class of preferences satisfying the following definition:

\begin{definition}
	A monotone ordinal preference $\succeq$ is \emph{responsive}
	if there exists a preference $\succeq^*$ over items such that $\forall S, \forall j,j'\notin S$, it holds that $S\cup\{j\}\succeq S\cup\{j'\} \iff j\succeq^*j'$ and $S\cup\{j\}\succ S\cup\{j'\} \iff j\succ^*j'$. 	
\end{definition}

Intuitively, an agent with a responsive preference would always trade a lower-ranked item for a higher-ranked one. Responsive preferences arise naturally in the matching literature. For example, consider a hospital that ranks individual doctors based on their test scores, then the hospital's preference over sets of doctors will be responsive to its preference over individuals.

\begin{proposition}
	\label{pro:classes}
	The strict inclusion relations in \eqref{eq:hierarchy} hold.
\end{proposition}

\proof 
For simplicity of exposition we show the inclusions for strict preferences; indifferences do not change the relations.
\begin{itemize}
	\item $\LEX \subsetneq\ADD$.
	Let $\prec$ be any preference in $\LEX$ induced by a preference $\prec^*$ over items. Wlog assume $1\prec^*2\prec^*\dots\prec^*m$. 
	Define an additive valuation $v$ in which item $j$ has value $2^{j-1}$ (larger than the total value of all items ranked lower than $j$). Observe that $v$ represents the preference $\prec$. Thus $\LEX \subseteq\ADD$, and
	it is not hard to see there are preferences in $\ADD\setminus\LEX$. 
	
	\item $\ADD \subsetneq\RSPN$.
	It is not hard to see that $\ADD \subseteq\RSPN$; we now construct a preference in $\RSPN\setminus\ADD$.
	Let~$\succ$ be a preference in $\RSPN$ induced by the preference $A\succ^*B\succ^*C\succ^*D\succ^*E$ over items.
	The preference $\succ^*$ over items induces a partial order over bundles of the same cardinality. For bundles of size 2, we know that either $AB\succ AC\succ BC\succ AD\succ BD\succ CD\succ AE\succ BE\succ CE\succ DE$, or $AB\succ AC\succ BC\succ AD\succ BD\succ AE\succ CD\succ BE\succ CE\succ DE$. However, $\succ^*$ does not induce an order over bundles of different cardinalities, which may be arbitrary subject to monotonicity. We may thus assume that both $AD \succ  BCD$ and $BCE \succ  AE$, without violating the responsiveness or monotonicity of preference~$\succ$. Now assume for contradiction that preference $\succ$ is represented by an additive valuation $v$. Then $AD \succ  BCD \implies v(A)>v(BC)$, but $BCE \succ AE \implies v(BC)>v(A)$, contradiction.  
	
	\item $\RSPN \subsetneq\SUBMODULAR\subsetneq\GEN$. It is not hard to see there are preferences in $\GEN\setminus\SUBMODULAR$ and in $\SUBMODULAR\setminus\RSPN$.
\end{itemize}
\endproof

\subsubsection{Submodular preferences.}

A well-known condition on preferences required by general equilibrium theory is \emph{non-satiation}: for every set $T$ and item $j\notin T$, it holds that
$T \prec T\cup \{j\}$ (stated otherwise, every set is strictly preferred over any of its subsets). For example, a \emph{strict} monotone preference clearly satisfies non-satiation.
We now weaken this condition: a monotone preference $\preceq_i$ satisfies \emph{satiation-only-at-the-top} when for every set $T\subset S$ and item $j\notin S$, 
if $T=_i T \cup\{j\}$ then $S=_i S \cup\{j\}$. That is, if given a set $T$, agent $i$ gains nothing by adding item $j\notin T$, then she also gains nothing by adding item $j$ to any superset of $T$ (if the preference is satiated for item $j$ at set $T$ then $T$ is ``at the top''). 
We denote by $\SATATTOP$ the class of preferences that satisfy the weakened condition.
One example of such preferences is those arising from unit-demand valuations,%
\footnote{A unit-demand valuation assigns a value to each item, and values every bundle by the highest-valued item within it.} 
or similarly from strict preferences over items but with several substitutable copies of every item.
As another example, consider the preference represented by valuation 
$v(\emptyset)=0<v(\{A\})=v(\{B\})=1<v(\{A,B\})=10$. 
While items $A,B$ are complements in this valuation, it satisfies non-satiation and thus also satiation-only-at the-top. 

\begin{theorem}
	\label{thm:submod-is-gen}
	Every monotone preference $\preceq$ that satisfies satiation-only-at-the-top (or non-satiation) can be represented by a submodular valuation; that is, $\SUBMODULAR = \SATATTOP$.
\end{theorem}

\proof
Index the bundles such that $\emptyset \preceq S_1 \preceq S_2 \preceq \dots \preceq S_{2^m-1}$ (with any order between bundles for which there is indifference). 
The \emph{layer} $l(t)$ of a set $S_t$ is defined as follows:  If $S_t=\emptyset$ then its layer is $0$. Otherwise it is at layer $k>0$ if there exists a set $T\prec S_t$ such that $T$ has layer $k-1$, and there is no set $X$ such that $S_t\prec X\prec T$.  
Define $v(S_t)=1-2^{-l(t)}$. 
It is not hard to see that $v$ represents $\preceq$. 

We now argue that $v$ is submodular, by showing that for every two sets $S_t\subset S_{t'}$ and item $j\notin S_{t'}$ it holds that $v(j\mid S_{t})\geq v(j\mid S_{t'})$. By monotonicity it holds that $t'>t$, $v(j\mid S_{t})\geq 0$ and $v(j\mid S_{t'})\geq 0$.
If $v(j\mid S_{t'})=0$ then clearly $v(j\mid S_{t})\geq 0 = v(j\mid S_{t'})$ as needed. So it remains to handle the case that $v(j\mid S_{t'})>0$. 
If $l(t)=l(t')$ then $v(S_{t'}) = v(S_{t})$ and thus $v(j\mid S_{t'}) = v(S_{t'}\cup \{j\}) - v(S_{t'})\geq  v(S_{t}\cup \{j\}) - v(S_{t})= v(j\mid S_{t}) $ by monotonicity. Otherwise, by monotonicity, $l(t')\geq l(t)+1$.

Since $v(S_{t'}) = 1 - 2^{-l(t')}$ and $v(S_{t'}\cup \{j\})<1$ we have 
$v(j\mid S_{t'}) = v(S_{t'}\cup \{j\}) - v(S_{t'}) < 1 - (1 - 2^{-l(t')})= 2^{-l(t')} \leq 2^{-(l(t)+1)}$, as $l(t')\geq l(t)+1$.

Additionally, by satiation-only-at-the-top, as $v(j\mid S_{t'})>0$ it must be the case that $v(j\mid S_{t})>0$.
Since $v(S_{t}) \prec v(S_{t}\cup \{j\})$, the set $S_{t}\cup \{j\}$ is at layer at least $l(t)+1$. Thus 
$ v(j\mid S_t) = v(S_t\cup \{j\}) - v(S_t) \ge 2^{-l(t)}-2^{-(l(t)+1)} = 2^{-(l(t)+1)}\geq v(j\mid S_{t'})$, completing the proof.
\Halmos
\endproof

We observe that the theorem is tight, thus providing a complete characterization of monotone preferences that can be represented by submodular valuations: if a monotone ordinal preference violates satiation-only-at-the-top, then item $j$ must have a positive marginal with respect to some set and a zero marginal with respect to its subset, violating submodularity. 

\subsubsection{Lexicographic preferences.}

\begin{proposition}
	For $n$ agents with lexicographic preferences and distinct budgets, a CE exists.
\end{proposition}

\proof
Run the well-known \emph{serial dictatorship} allocation protocol (see \cite{AS98}) as follows: Order the agents by decreasing budget and let each agent $i<n$ in turn pick her most preferred item among the unallocated items. Price this item at $b_i$ (agent $i$'s budget). Allocate all leftover items to agent $n$ whose budget is smallest, pricing such items at $b_n/r$ where $r$ is the number of leftovers. The resulting allocation and item pricing form a CE.%
\endproof

	\section{Summary}
	\label{sec:summary}
		We view our main conceptual contribution as showing that to get positive equilibrium existence results (and associated fairness guarantees) for interesting classes of markets beyond additive preferences, it is sufficient to exclude degenerate market instances. 
	This is done by adding ``small'' noise to the agents' budgets to make them generic, in the spirit of \emph{smoothed analysis} for getting positive computational tractability results \cite{ST09}. Unlike previous approaches, there is no need to relax the equilibrium notion, and our model allows for possibly very different budgets. Our approach is shown to work for markets with general ordinal preferences and a small number of items, markets with 2 agents with leveled preferences, and markets with $n$ agents with lexicographic preferences. We expect additional classes of markets to be added to this list in the future. Towards this end, we explore a hierarchy of preference classes inspired by the combinatorial auctions literature, and present a negative result for the highest level of the hierarchy. Further mapping of the theoretical landscape (for example, exploring the case of symmetric preferences) remains a promising direction for future research, as well as understanding why and to what extent generic CE existence is common in practice. 

	\bibliographystyle{named}
	\bibliography{abb,MOR-bib}

\begin{thebibliography}{}

\bibitem[\protect\citeauthoryear{Abdulkadiro\u{g}lu and S\"onmez}{1998}]{AS98}
Abdulkadiro\u{g}lu and S\"onmez.
\newblock Random serial dictatorship and the core from random endowments in
  house allocation problems.
\newblock {\em Econometrica}, 66(3):689--701, 1998.

\bibitem[\protect\citeauthoryear{Arrow and Debreu}{1954}]{AD54}
Kenneth~J. Arrow and Gerard Debreu.
\newblock Existence of equilibrium for a competitive economy.
\newblock {\em Econometrica}, 22(3):265--290, 1954.

\bibitem[\protect\citeauthoryear{Babaioff \bgroup \em et al.\egroup
  }{2019}]{BabaioffNT17}
Moshe Babaioff, Noam Nisan, and Inbal Talgam{-}Cohen.
\newblock Fair allocation through competitive equilibrium from generic incomes.
\newblock In {\em Proceedings of the 2nd ACM Conference on Fairness,
  Accountability, and Transparency}, pages 180--180, 2019.

\bibitem[\protect\citeauthoryear{Barman \bgroup \em et al.\egroup
  }{2018}]{BarmanKV18}
Barman, Krishnamurthy, and Vaish.
\newblock Finding fair and efficient allocations.
\newblock In {\em Proceedings of the 19th EC}, pages 557--574, 2018.

\bibitem[\protect\citeauthoryear{Blumrosen and Nisan}{2007}]{BN07}
Liad Blumrosen and Noam Nisan.
\newblock Combinatorial auctions.
\newblock In {\em Algorithmic Game Theory}, chapter~11. Cambridge University
  Press, 2007.

\bibitem[\protect\citeauthoryear{Bouveret \bgroup \em et al.\egroup
  }{2016}]{BCM16}
Sylvain Bouveret, Yann Chevaleyre, and Nicolas Maudet.
\newblock Fair allocation of indivisible goods.
\newblock In {\em Handbook of Computational Social Choice}, chapter~12, pages
  284--310. Cambridge University Press, 2016.

\bibitem[\protect\citeauthoryear{Brainard and Scarf}{2005}]{BS05}
William~C. Brainard and Herbert~E. Scarf.
\newblock How to compute equilibrium prices in 1981.
\newblock {\em The American Journal of Economics and Sociology}, 64(1):57--83,
  2005.

\bibitem[\protect\citeauthoryear{Brams and Kaplan}{2004}]{BK04}
Steven~J. Brams and Todd~R. Kaplan.
\newblock Dividing the indivisible.
\newblock {\em Journal of Theoretical Politics}, 16(2):143--173, 2004.

\bibitem[\protect\citeauthoryear{Brams and Taylor}{1996}]{BT96}
Steven~J. Brams and Alan~D. Taylor.
\newblock {\em Fair Division: From Cake-Cutting to Dispute Resolution}.
\newblock Cambridge University Press, 1996.

\bibitem[\protect\citeauthoryear{Brandt \bgroup \em et al.\egroup
  }{2016}]{BCE+16}
Felix Brandt, Vincent Conitzer, Ulle Endriss, Jerome Lang, and Ariel~D.
  Procaccia.
\newblock {\em Handbook of Computational Social Choice}.
\newblock Cambridge University Press, 2016.

\bibitem[\protect\citeauthoryear{Br\^anzei \bgroup \em et al.\egroup
  }{2015}]{BHM15}
Simina Br\^anzei, Hadi Hosseini, and Peter~Bro Miltersen.
\newblock Characterization and computation of equilibria for indivisible goods.
\newblock In {\em Proceedings of the 8th SAGT}, pages 244--255, 2015.

\bibitem[\protect\citeauthoryear{Budish}{2011}]{Bud11}
Eric Budish.
\newblock The combinatorial assignment problem.
\newblock {\em Journal of Political Economy}, 119(6):1061--1103, 2011.

\bibitem[\protect\citeauthoryear{Cramton \bgroup \em et al.\egroup
  }{1987}]{CGK87}
Peter Cramton, Robert Gibbons, and Paul Klemperer.
\newblock Dissolving a partnership efficiently.
\newblock {\em Econometrica}, 55(3):615--632, 1987.

\bibitem[\protect\citeauthoryear{Deng \bgroup \em et al.\egroup }{2003}]{DPS03}
Xiaotie Deng, Christos~H. Papadimitriou, and Shmuel Safra.
\newblock On the complexity of price equilibria.
\newblock {\em J. Comput. Syst. Sci.}, 67(2):311--324, 2003.

\bibitem[\protect\citeauthoryear{Farhadi \bgroup \em et al.\egroup
  }{2017}]{FGH+17}
Farhadi, Ghodsi, Hajiaghayi, Lahaie, Pennock, Seddighin, Seddighin, and Yami.
\newblock Fair allocation of indivisible goods to asymmetric agents.
\newblock In {\em Proceedings of the 16th AAMAS}, pages 1535--1537, 2017.

\bibitem[\protect\citeauthoryear{Ghodsi \bgroup \em et al.\egroup
  }{2011}]{GZH+11}
Ali Ghodsi, Matei Zaharia, Benjamin Hindman, Andy Konwinski, Scott Shenker, and
  Ion Stoica.
\newblock Dominant resource fairness.
\newblock In {\em Proceedings of the 8th NSDI}, pages 305--322, 2011.

\bibitem[\protect\citeauthoryear{Goldman and Procaccia}{2014}]{GP14}
Jonathan Goldman and Ariel~D. Procaccia.
\newblock Spliddit.
\newblock {\em ACM SIGecom Exchanges}, 13(2):41--46, 2014.

\bibitem[\protect\citeauthoryear{Gul and Stacchetti}{1999}]{GS99}
Faruk Gul and Ennio Stacchetti.
\newblock Walrasian equilibrium with gross substitutes.
\newblock {\em Journal of Economic Theory}, 87:95--124, 1999.

\bibitem[\protect\citeauthoryear{Lehmann \bgroup \em et al.\egroup
  }{2006}]{LLN06}
Benny Lehmann, Daniel Lehmann, and Noam Nisan.
\newblock Combinatorial auctions with decreasing marginal utilities.
\newblock {\em Games and Economic Behavior}, 55:270--296, 2006.

\bibitem[\protect\citeauthoryear{Mas-Colell}{1977}]{Mas77}
Andreu Mas-Colell.
\newblock Indivisible commodities and general equilibrium theory.
\newblock {\em Journal of Economic Theory}, 16(2):443--456, 1977.

\bibitem[\protect\citeauthoryear{Maskin}{1987}]{Mas87}
Eric~S. Maskin.
\newblock On the fair allocation of indivisible goods.
\newblock In G.~Feiwel, editor, {\em Arrow and the Foundations of the Theory of
  Economic Policy}, pages 341--349. MacMillan Publishing Company, 1987.

\bibitem[\protect\citeauthoryear{Niazadeh and Wilkens}{2016}]{NW16}
Rad Niazadeh and Christopher Wilkens.
\newblock Competitive equilibria for non-quasilinear bidders in combinatorial
  auctions.
\newblock In {\em Proceedings of the 12th WINE}, pages 116--130, 2016.

\bibitem[\protect\citeauthoryear{Othman \bgroup \em et al.\egroup
  }{2016}]{OPR16}
Abraham Othman, Christos~H. Papadimitriou, and Aviad Rubinstein.
\newblock The complexity of fairness through equilibrium.
\newblock {\em ACM Trans. Economics and Comput.}, 4(4):20, 2016.

\bibitem[\protect\citeauthoryear{Pratt and Zeckhauser}{1990}]{PZ90}
Pratt and Zeckhauser.
\newblock The fair and efficient division of the {W}insor family silver.
\newblock {\em Management Science}, 36(11):1293--1301, 1990.

\bibitem[\protect\citeauthoryear{Rastogi and Cole}{2007}]{RC07}
Ashish Rastogi and Richard Cole.
\newblock Indivisible markets with good approximate equilibrium prices.
\newblock ECCC, 2007.

\bibitem[\protect\citeauthoryear{Robertson and Webb}{1998}]{RW98}
Jack Robertson and William Webb.
\newblock {\em Cake-Cutting Algorithms}.
\newblock Peters/CRC Press, 1998.

\bibitem[\protect\citeauthoryear{Segal-Halevi}{2018}]{Seg18}
Erel Segal-Halevi.
\newblock Competitive equilibrium for almost all incomes.
\newblock In {\em Proceedings of the 17th AAMAS}, pages 1267--1275, 2018.

\bibitem[\protect\citeauthoryear{Shapley and Scarf}{1974}]{SS74}
Lloyd Shapley and Herbert Scarf.
\newblock On cores and indivisibility.
\newblock {\em Journal of Mathematical Economics}, 1(1):23--37, 1974.

\bibitem[\protect\citeauthoryear{Spielman and Teng}{2009}]{ST09}
Daniel~A. Spielman and Shang-Hua Teng.
\newblock Smoothed analysis.
\newblock {\em Communications of the ACM}, 52(10):76--84, 2009.

\bibitem[\protect\citeauthoryear{Svensson}{1983}]{Sve83}
Lars-Gunnar Svensson.
\newblock Large indivisibles: An analysis with respect to price equilibrium and
  fairness.
\newblock {\em Econometrica}, 51(4):939--954, 1983.

\bibitem[\protect\citeauthoryear{Svensson}{1984}]{Sve84}
Lars-Gunnar Svensson.
\newblock Competitive equilibria with indivisible goods.
\newblock {\em Journal of Economics}, 44(4):373--386, 1984.

\end{thebibliography}
		
\end{document}